\documentclass[aps,prl,twocolumn,superscriptaddress,showpacs]{revtex4}

\usepackage{graphicx}
\begin{document}

\title{Spin Susceptibility of an Ultra-Low Density Two Dimensional Electron System}

\author{J. Zhu}
\affiliation{Department of Physics, Columbia University, New York, New York 10027}

\author{H. L. Stormer}
\affiliation{Department of Physics, Columbia University, New York, New York
10027}\affiliation{Department of Applied Physics and Applied Mathematics, Columbia
University, New York, New York 10027} \affiliation{Bell Labs, Lucent Technologies, Murray
Hill, New Jersey 07974}
\author{L. N. Pfeiffer}
\affiliation{Bell Labs, Lucent Technologies, Murray Hill, New Jersey 07974}
\author{K. W. Baldwin}
\affiliation{Bell Labs, Lucent Technologies, Murray Hill, New Jersey 07974}
\author{K. W. West}
\affiliation{Bell Labs, Lucent Technologies, Murray Hill, New Jersey 07974}

\date{December 10, 2002}

\begin{abstract}
We determine the spin susceptibility in a two dimensional electron system in
GaAs/AlGaAs over a wide range of low densities from 2$\times10^{9}$cm$^{-2}$ to
4$\times10^{10}$cm$^{-2}$. Our data can be fitted to an equation that describes
the density dependence as well as the polarization dependence of the spin
susceptibility. It can account for the anomalous g-factors reported recently in
GaAs electron and hole systems. The paramagnetic spin susceptibility increases
with decreasing density as expected from theoretical calculations.
\end{abstract}
\pacs{73.40.-c, 71.27+a, 71.30+h}
 \maketitle

The low-density ground state of a degenerate electron system is
one of the oldest questions of many particle physics. In two
dimensions, it is expected that in the dilute limit, the electron
Fermi liquid undergoes a phase transition to a solid known as the
Wigner
crystal\cite{Wigner34crystal,Chaplik72crystal,Tanatar89crystal}.
Furthermore, driven by the exchange coupling between electrons, a
ferromagnetic state may arise at densities slightly above the
critical density of Wigner
crystallization\cite{Bloch29ferro,Stoner38ferro,Varsano01ferro,Moroni02ferro}.
None of these phases have yet been detected due to the lack of
sufficiently high-quality, low-density specimens. Recent
measurements of relevant system parameters such as the spin
susceptibility $\chi$ and the effective g-factor g$^{*}$ have
shown considerable deviations from their standard, non-interacting
values\cite{Okamoto99bparal,Shashkin01silicon,Pudalov02silicon,Tutuc02gfactor,Noh02gfactor}.
This is believed to be the result of strong
interaction\cite{Varsano01ferro,Moroni02ferro}, but disorder-led
enhancement cannot be ruled out. Such explorations of the
pre-transition regime provide valuable data against which to check
the same theoretical calculations that determine the transition
point to either a ferromagnetic state or to the Wigner solid. A
possible connection between the ferromagnetic properties and the
metal-to-insulator transition (MIT) in low density 2DESs has added
to the complexity of the problem and prompted intense recent
studies on this subject in Silicon
MOSFETs\cite{Rev01MIT,Vitkalov01silicon,Shashkin01silicon,Pudalov02silicon,Shashkin02mass}.
These results generally lend support to the possibility of a
ferromagnetic state. On the other hand two groups have recently
reported anomalous density dependencies of the g-factors in the
GaAs/AlGaAs electron and hole systems that are at odds with
results in Silicon MOSFETs and disfavor the predicted
ferromagnetic transition\cite{Tutuc02gfactor,Noh02gfactor}.

In this paper, we report measurements of $\chi$ in a variable density 2DES of
exceedingly high quality to unprecedented low densities. In addition to the
density dependence of $\chi$, our measurements determine, for the first time,
the explicit polarization dependence of $\chi$. This polarization dependence
can account for the `` anomalous'' g-factors recently reported in GaAs 2DESs
and 2D hole systems (2DHG)\cite{Tutuc02gfactor,Noh02gfactor}.

Our specimen is a heterojunction-insulated gate field effect transistor
(HIGFET). The specimen consists of a (001) GaAs substrate, overgrown by
molecular beam epitaxy with 0.5$\mu$m of GaAs, followed by a 200-fold
superlattice of 10nm of GaAs and 3nm of Al$_{0.32}$Ga$_{0.68}$As. Subsequently,
2$\mu$m of GaAs are deposited as a channel, followed by 600nm of
Al$_{0.32}$Ga$_{0.68}$As as an effective insulator, and capped by a heavily
doped GaAs n$^{+}$ layer, serving as a top gate. The specimen is processed into
a 600 $\mu$m square mesa. Sixteen Ni-Ge-Au contact pads are spaced evenly along
the edges of the mesa using standard photolithography. One corner pad provides
the contact to the top gate, which allows for a continuous and \textit{in situ}
change of the 2DES density. The density range available for measurement extends
from 1.7x10$^{9}$cm$^{2}$ to 6.4x10$^{10}$cm$^{2}$. The interaction parameter
r$_{s}$, defined as the ratio of the inter-electron spacing to the Bohr radius,
a/a$_{B}$=1/$\sqrt{\pi n}$a$_{B}$, spans 2.2 $<$ r$_{s}$ $<$ 13.4. Screening of
the e-e interaction via the top metallic gate can be neglected, since its
distance exceeds the electron spacing by a factor of 4.4 to 27 in our
experiments. The MIT occurs at 2x10$^{9}$cm$^{2}$, which is the lowest
transition density ever reported in a 2D system, attesting to the low disorder
of the specimen. Our FET gives us the unique advantage of wide density
tunability within one specimen and its reproducible gate-voltage/density
relation lets us sweep density at fixed magnetic field, which is an essential
method in our $\chi$ determination.

Our measurements were performed in a dilution refrigerator equipped with a rotating
sample platform reaching a base temperature of 30mK. A standard low frequency (3-23Hz)
lock-in technique was used with excitation currents ranging from 100pA to 100nA. All
experiments were performed during one cooldown, facilitating quantitative comparison
between our data.

In a normal Fermi liquid, the spin susceptibility $\chi$ =d$\Delta$n/dB=
g$^{*}\mu_{B}\rho/2$, where g$^{*}$ is the effective g-factor, $\rho$ is the
density of states (DOS) at the Fermi level, and
$\Delta$n=n$^{\uparrow}$-n$^{\downarrow}$. In a 2D system,
$\rho=m^{*}/\pi\hbar^{2}$, therefore $\chi=g^{*}m^{*}\mu_{B}/2\pi\hbar^{2}$.
Generally, theories find that $\chi$ increases with growing interaction due to
spin-exchange coupling. To investigate the experimental situation, we measure
the spin susceptibility $\chi$ as a function of the 2DES density n. We express
it as a relative spin susceptibility $\chi/\chi_{0}= m^{*}g^{*}/m_{b}g_{b}$,
where m$_{b}$=0.067m$_{e}$ and g$_{b}$=0.44 are the band values of mass and
g-factor in GaAs and $\chi_{0}$ the Pauli susceptibility determined by these
band values. In the remainder of the paper, we use values for $\chi$ and m*g*
normalized to $\chi_{0}$ and m$_{b}$g$_{b}$ respectively, so that $\chi$= m*g*.
We employ two different methods to measure m*g*. First, we follow and extend
the tilted field method introduced by Fang and Stiles in Silicon
MOSFET\cite{Fang68tilt}. Secondly, we follow the parallel-field method recently
utilized by Refs.~\cite{Tutuc02gfactor,Noh02gfactor} to derive m*g* from the
full polarization condition of the 2DES.

\begin{figure}
\includegraphics{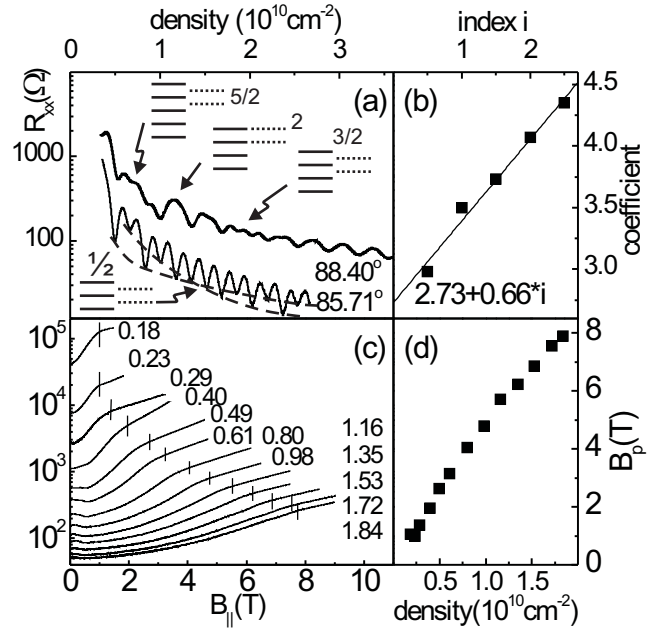}
 \caption{\label{trace}(a) Density sweeps at $B_{perp}$=0.07T and $\theta$
= 85.71$^{\circ}$ (thin) and 88.40$^{\circ}$ (thick) respectively. Events corresponding
to g*$\mu_{B}$B$_{tot}$=i$\hbar\omega_{c}$ where i=1/2, 3/2, 2 and 5/2 are indicated in
the plot with level diagrams nearby as insets. Dashed lines illustrate the procedure to
determine n$_{0}$ at half-integer configurations. (b) Linear dependence of coefficients
of the power law fits of Fig.~\ref{susceptibility} on their index, i. (c)
Magneto-resistances in parallel-field method for different densities in unit of
10$^{10}$cm$^{-2}$. Positions of full polarization fields, B$_{p}$, are indicated
following Ref.\cite{Tutuc02gfactor}. (d) B$_{p}$ as a function of density.}
\end{figure}

In a magnetic field, spin-up and spin-down electrons form two separate sets of
landau levels. As the magnetic field is tilted, the two sets of landau levels
move with respect to each other. The energy diagrams are schematically shown as
insets in Fig.~\ref{trace}(a). Solid and dotted lines represent spin-up and
spin-down Landau levels, respectively. The spacing between Landau levels,
$\hbar\omega_{c}=e\hbar$B$_{perp}$/m*, depends on the perpendicular component
of the field. The shift between both sets, on the other hand, is the Zeeman
energy, $\Delta E_{z}=g^{*}\mu_{B}$B$_{tot}$, which depends on the total field.
By adjusting the tilt angle, $\theta$ and the total field, B$_{tot}$,
$\hbar\omega_{c}$ and $\Delta E_{z}$ can be independently changed. Particularly
useful configurations arise when g$^{*}\mu_{B}$B$_{tot}$=i$\hbar\omega_{c}$,
where i is an integer or a half-integer. At half-integer configurations, the
Landau levels from both spins interleave and form a set of uniformly spaced
levels. At integer configurations the Landau levels coincide and form again a
set of uniformly spaced levels, however, this time with double spacing compared
to the half-integer case (see insets Fig.~\ref{trace}(a)).

\begin{figure}
\includegraphics{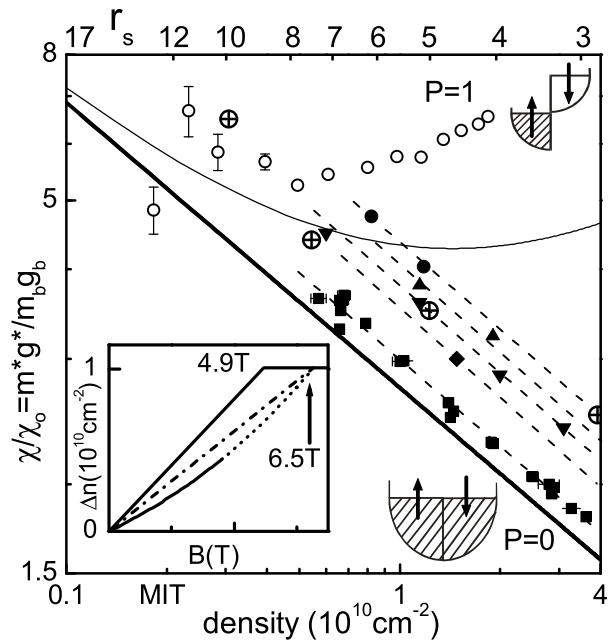}

 \caption{\label{susceptibility}Density dependence of m*g* determined by
two different methods. Solid data points are from tilted-field experiment with
different indices: square for i=1/2, diamond for i=1, down triangle for i=3/2,
up triangle for i= 2, circle for i=5/2. Parallel dashed lines indicate power
law dependence of m*g* with a single exponent for all i. Their coefficients
depend linearly on i, see Fig.~\ref{trace}(b). Open circles show non-monotonic
density dependence of m*g* derived from full polarization field, B$_{p}$, of
in-plane field method. Inset shows net spin $\Delta$n=Pn for
n=1x10$^{10}$cm$^{-2}$ with interpolated regime (solid line) and extrapolated
regime (dotted line). B$_{p}$=4.9T from in-plane field method and
B$_{ext}$=6.5T from extrapolation of tilted-field method. A nominal
m*g*$_{ext}$, slope of the dash-dotted line is derived from B$_{ext}$ for all
densities and plotted as a thin solid line in full figure. Thick solid line
represents extrapolation of m*g* to P=0 limit. Calculations from
Ref.~\cite{Moroni02ferro} are shown as crossed circles.}
\end{figure}

The magneto-resistance of the 2DES reflects the configuration of the energy
levels. Distinctive signatures from different configurations are observed when
the 2DES density is swept at fixed B$_{tot}$ and $\theta$. B$_{perp}$, and
hence $\theta$, is accurately determined from the period of the oscillations
and the Landau level degeneracy eB$_{perp}$/h. At half-integer configurations,
the depths of successive minima are just equal. This is also the case for the
integer configurations, but here each second minimum disappears. The different
configurations are realized only transiently at particular densities n$_{0}$
that satisfy g$^{*}\mu_{B}$B$_{tot}$=i$\hbar\omega_{c}$, equivalent to
m*g*(n$_{0}$)=ie$\hbar\cos{\theta}/\mu_{B}$, since m*g* depends on density.
Integer configurations are easily identified by the disappearance of a minimum.
At half-integer configurations, the depth of neighboring minima interchange
strength at n$_{0}$, which is identified by the crossing point of two smooth
envelopes drawn along alternate minima, see Fig.~\ref{trace}(a) for
85.71$^{\circ}$. The trace at 88.40$^{\circ}$ of Fig.~\ref{trace}(a) shows, as
examples, several configurations as the density is swept and indicated by the
neighboring diagrams. Increasing $\theta$ slowly and tracking the indices
carefully, we are able to identify and label events that belong to
configurations with index 1/2, 1, 3/2, 2, and 5/2. The product m*g* is
calculated according to m*g*=ie$\hbar\cos{\theta}/\mu_{B}$. The solid symbols
in Fig.~\ref{susceptibility} represent data derived with this method, using
different symbols for different indices.

Clearly, m*g* increases with decreasing density and increasing index i. More
strikingly, for each fixed index i, m*g* displays a power law dependence on n
within the measured range, indicated by the dashed lines on a log-log scale.
This dependence is best developed for i=1/2 and 3/2. Furthermore, the parallel
lines suggest a single exponent. We therefore fit a power law dependence to the
data of i=1/2 and use the exponent as a constraint in fitting data of the other
indices. The coefficients from these five fits display a very good linear
dependence on i, as seen in Fig.~\ref{trace}(b). This gives us an empirical
equation, m*g* =(2.73+0.66*i)n$^{-0.4}$, where n is in unit of
10$^{10}$cm$^{-2}$. This equation describes all our data points remarkably
well.

Before discussing the implications of such an empirical equation, we proceed
with the second method of determining m*g*. With increasing in-plane magnetic
field, B$_{\parallel}$, the polarization, P=(n$^{\uparrow}$-n$^{\downarrow}$)/n
of the 2DES increases and saturates at unity at a threshold field $B_{p}$ when
g*$\mu_{B}$B$_{p}$/2=E$_{F}$. The derivative of $\Delta$n (=Pn) with respect to
B$_{\parallel}$ is $\chi$(= m*g*). Assuming m*g* to be independent of P it
follows that m*g*=2$\pi\hbar^{2}$n/$\mu_{B}$B$_{p}$. As asserted by several
groups, full polarization of the 2DES is signaled by the onset of an
exponential behavior in a parallel-field magneto-resistance
experiment\cite{Okamoto99bparal,Vitkalov00bparal,Dolgopolov00bparal,Tutuc01bparal,Herbut01inplanefield}
We have performed such experiments on our HIGFET for different densities
(Fig.~\ref{trace}(c)), determined B$_{p}$ according to
Ref.~\cite{Tutuc02gfactor}, and show their values in Fig.~\ref{trace}(d).
Within experimental error, B$_{p}$ is independent of the current direction
relative to B$_{\parallel}$. Translated into m*g* the results are plotted as
open circles in Fig.~\ref{susceptibility}. Unlike the solid data points from
the first determination of m*g*, these new data exhibit a non-monotonic
dependence on density and are consistently larger in value than the solid data
points, with the discrepancy increasing with increasing density.

The key to reconciling the discrepancy between these data sets is to recognize
the dependence of m*g* on the polarization, P. In the tilted field method, a
higher index i implies a higher Zeeman energy and therefore a higher degree of
2DES polarization. In the Fermi liquid limit (B$_{perp}$=0),
P=(n$^{\uparrow}$-n$^{\downarrow}$)/n = $\Delta
E_{z}\rho$/2n=(g*$\mu_{B}$B$_{tot}$)(m*/2$\pi\hbar^{2}$)/n. The tilted field
method assumes that the introduction of a small B$_{perp}$ does not alter the
values of g* and m*. Using the relation
g*$\mu_{B}$B$_{tot}$=i$\hbar\omega_{c}$, a straightforward calculation yields
P=ieB$_{perp}$/nh. This direct relationship between P and index i establishes
an explicit P-dependence of m*g*. Substituting P for i and using
B$_{perp}$=0.07T\cite{note0.05T}, we obtain $\chi$=m*g*=(2.73+3.9Pn)n$^{-0.4}$
from our data. Most remarkably, for fixed density n, the susceptibility $\chi$
increases linearly with increasing polarization. This monotonic increase
reflects an increasing spin-exchange energy with increasing population of
like-spins.

Quite clearly, the assumption of a P-independent $\chi$, assumed in
Refs.~\cite{Tutuc02gfactor,Noh02gfactor} to derive m*g* from B$_{p}$, does not
apply. Our empirical interpolation equation can be used to make contact between
the data obtained from the tilted-field method and from the parallel field
method. Their relationship is best discussed referring to the inset of
Fig.~\ref{susceptibility}. This diagram shows, as an example, the evolution of
the net spin $\Delta$n=Pn=n$^{\uparrow}$-n$^{\downarrow}$ with B$_{tot}$ for a
fixed density, n=1x10$^{10}$cm$^{-2}$. By definition, the slope of $\Delta$n(B)
is the susceptibility $\chi$= m*g*. The parallel-field method determines
$\Delta$n=n (=1x10$^{10}$cm$^{-2}$) at B$_{p}$=4.9T. The assumption of a
B-independent m*g* implies a linear rise of $\Delta$n with B, shown as a
straight line in the inset. One value, B$_{p}$=4.9T, determines m*g*$_{p}$=
2$\pi\hbar^{2}$n/$\mu_{B}$B$_{p}$=5.7. However, the relationship between
$\Delta$n and B is not linear but must be deduced from the tilted-field
experiments. We derive the actual $\Delta$n vs B curve by integrating the
empirical equation of $\chi$(P) and obtain
$\Delta$n=0.69[$\exp(0.138Bn^{-0.4})-1]$. The curve for n=1x10$^{10}$cm$^{-2}$
is also shown in the inset. The solid portion represents the interpolated
regime of $\chi$ while the dotted portion represents extrapolation of $\chi$
beyond our tilted field data. Requiring $\Delta$n=n (=1x10$^{10}$cm$^{-2}$), we
obtain $B_{ext}$=6.5T for the full polarization field, which is 33\% higher
than the measured value of B$_{p}$. This B$_{ext}$ yields a "nominal"
m*g*$_{ext}$ = 2$\pi\hbar^{2}$n/$\mu_{B}$B$_{ext}$ =4.3, equivalent to the
slope of the dash-dotted line in the inset.

Performing this derivation for all densities using the interpolation equation
we arrive at an m*g*$_{ext}$, which is plotted as a thin, curved line in
Fig.~\ref{susceptibility}. It shows a qualitative similarity to the m*g*$_{p}$
data derived from the parallel field measurements, particularly the
non-monotonic density dependence. In fact, in the low density limit our derived
curve matches m*g*$_{p}$ very well, indicating that our empirical equation
extrapolates well into this regime. At higher densities considerable
discrepancies arise, which one may, at first glance, attribute to a
discontinuity in $\Delta$n on the polarization curve, indicating the existence
of a first order phase transition. However, given the good agreement between
m*g*$_{p}$ and m*g*$_{ext}$ at low densities, and hence the absence of a phase
transition in this regime, it is unlikely that such a transition occurs at
higher densities. Instead, we suspect that the extrapolation of our empirical
equation becomes less accurate with increasing density due to the increasing
range of extrapolation and the solid curve of the inset bends sharply, but
continuously towards B=4.9T. A mass increase, caused by the in-plane field,
could provide a mechanism for such an accelerated bending. This effect is
negligible at low field and hence low density, but increases rapidly with
B$_{\parallel}$\cite{Batke86massinplanefield,Batke88massinplanefield}.
Alternatively, at high polarization, terms of higher order in $\Delta$n may
increasingly contribute to $\chi$, leading to an accelerated bending as well.
In spite of this discrepancy for extrapolations at high densities, the
qualitative and partially quantitative agreement between m*g*$_{ext}$ and
m*g*$_{p}$ strongly suggests that our empirical equation captures correctly the
underlying physics of the system.

Our analysis of m*g* provides a simple interpretation of the non-monotonic behavior of
m*g* derived from the parallel-field data. The unusual density dependence of m*g*$_{p}$
results from a combination of the polarization and density dependence of $\chi$. Clearly,
this m*g* does \emph{not} reflect the actual susceptibility at any polarization value and
consequently its density dependence cannot be used to assess the possibility of a
ferromagnetic transition. By comparing measurement and extrapolation, we conclude that a
first order transition is unlikely to occur in our 2DES within the regime of r$_{s}$
studied (3-12.4). In the remainder of the paper, we examine $\chi$ in the limit of
vanishing polarization, which plays a central role in the context of a second order
ferromagnetic transition.

Our empirical formula $\chi$=m*g*=(2.73+3.9Pn)n$^{-0.4}$, is readily
extrapolated to P=0 to yield the spin susceptibility, $\chi$, of a normal Fermi
liquid (see Fig.~\ref{susceptibility}). From n=5x10$^9$cm$^{-2}$ to
4x10$^{10}$cm$^{-2}$, $\chi=2.73n^{-0.4}$ showing an enhancement factor of 1.6
to 3.6. This extrapolation is very reliable since it extends only slightly
beyond the range of our data. Furthermore, from the excellent agreement at low
densities between the m*g*$_{p}$ data and the extrapolated m*g*$_{ext}$ we are
confident that it provides very good estimates for $\chi$ to densities as low
as the MIT density n=2x10$^9$cm$^{-2}$ (r$_{s}$=12.4), where $\chi$ reaches
about 5.5. Whether a divergence occurs at yet lower density cannot be inferred
from our data.

Recent quantum Monte Carlo (QMC) calculations have predicted a weakly first
order ferromagnetic transition at r$_{s}$=20-30, corresponding to
3.4-7.6$\times10^8$cm$^{-2}$ in our 2DES\cite{Varsano01ferro,Moroni02ferro}.
This density range is currently beyond our reach. However, our measurement of
$\chi$ as the 2DES approaches the phase transition can be compared to theory,
thereby providing guidance as to the general applicability of the theoretical
approach. In Fig.~\ref{susceptibility}, the results of Ref.\cite{Moroni02ferro}
are plotted as large, crossed circles\cite{notemg}. The general trend of
increasing $\chi$ with increasing r$_{s}$ agrees very well with our data
although their estimates of $\chi$ seem to be offset to slightly larger values
compared to experiment.

In conclusion, we have determined the polarization and density dependence of
the enhanced spin susceptibility $\chi\sim$ m*g* in a 2DES in GaAs/AlGaAs over
a wide range of low densities. Our analysis provides an explanation for the
anomalous g-factors reported recently in GaAs 2DES and 2DHG. The susceptibility
$\chi$ of the unpolarized Fermi liquid increases with decreasing density, in
qualitative agreement with recent QMC calculations.

\begin{acknowledgments}
We thank D. C. Tsui for inspiring discussions and A. Millis for many insightful
suggestions. Financial support from the W. M. Keck Foundation is gratefully acknowledged.
\end{acknowledgments}


\begin{thebibliography}{24}
\expandafter\ifx\csname natexlab\endcsname\relax\def\natexlab#1{#1}\fi
\expandafter\ifx\csname bibnamefont\endcsname\relax
  \def\bibnamefont#1{#1}\fi
\expandafter\ifx\csname bibfnamefont\endcsname\relax
  \def\bibfnamefont#1{#1}\fi
\expandafter\ifx\csname citenamefont\endcsname\relax
  \def\citenamefont#1{#1}\fi
\expandafter\ifx\csname url\endcsname\relax
  \def\url#1{\texttt{#1}}\fi
\expandafter\ifx\csname urlprefix\endcsname\relax\def\urlprefix{URL }\fi
\providecommand{\bibinfo}[2]{#2} \providecommand{\eprint}[2][]{\url{#2}}

\bibitem[{\citenamefont{Wigner}(1934)}]{Wigner34crystal}
\bibinfo{author}{\bibfnamefont{E.}~\bibnamefont{Wigner}},
  \bibinfo{journal}{Phys. Rev.} \textbf{\bibinfo{volume}{46}},
  \bibinfo{pages}{1002} (\bibinfo{year}{1934}).

\bibitem[{\citenamefont{Chaplik}(1972)}]{Chaplik72crystal}
\bibinfo{author}{\bibfnamefont{A.~V.} \bibnamefont{Chaplik}},
  \bibinfo{journal}{Sov. Phys. Jetp.} \textbf{\bibinfo{volume}{35}},
  \bibinfo{pages}{395} (\bibinfo{year}{1972}).

\bibitem[{\citenamefont{Tanatar and Ceperley}(1989)}]{Tanatar89crystal}
\bibinfo{author}{\bibfnamefont{B.}~\bibnamefont{Tanatar}} \bibnamefont{and}
  \bibinfo{author}{\bibfnamefont{D.~M.} \bibnamefont{Ceperley}},
  \bibinfo{journal}{Phys. Rev. B} \textbf{\bibinfo{volume}{39}},
  \bibinfo{pages}{5005} (\bibinfo{year}{1989}).

\bibitem[{\citenamefont{Bloch}(1929)}]{Bloch29ferro}
\bibinfo{author}{\bibfnamefont{F.}~\bibnamefont{Bloch}}, \bibinfo{journal}{Z.
  Phys.} \textbf{\bibinfo{volume}{57}}, \bibinfo{pages}{545}
  (\bibinfo{year}{1929}).

\bibitem[{\citenamefont{Stoner}(1938)}]{Stoner38ferro}
\bibinfo{author}{\bibfnamefont{E.~C.} \bibnamefont{Stoner}},
  \bibinfo{journal}{Proc. R. Soc. London A} \textbf{\bibinfo{volume}{165}},
  \bibinfo{pages}{372} (\bibinfo{year}{1938}).

\bibitem[{\citenamefont{Varsano et~al.}(2001)\citenamefont{Varsano, Moroni, and
  Senatore}}]{Varsano01ferro}
\bibinfo{author}{\bibfnamefont{D.}~\bibnamefont{Varsano}},
  \bibinfo{author}{\bibfnamefont{S.}~\bibnamefont{Moroni}}, \bibnamefont{and}
  \bibinfo{author}{\bibfnamefont{G.}~\bibnamefont{Senatore}},
  \bibinfo{journal}{Europhys. Lett.} \textbf{\bibinfo{volume}{53}},
  \bibinfo{pages}{348} (\bibinfo{year}{2001}).

\bibitem[{\citenamefont{Attaccalite et~al.}(2002)\citenamefont{Attaccalite,
  Moroni, Gori-Giorgi, and Bachelet}}]{Moroni02ferro}
\bibinfo{author}{\bibfnamefont{C.}~\bibnamefont{Attaccalite}},
  \bibinfo{author}{\bibfnamefont{S.}~\bibnamefont{Moroni}},
  \bibinfo{author}{\bibfnamefont{P.}~\bibnamefont{Gori-Giorgi}},
  \bibnamefont{and} \bibinfo{author}{\bibfnamefont{G.~B.}
  \bibnamefont{Bachelet}}, \bibinfo{journal}{Phys. Rev. Lett.}
  \textbf{\bibinfo{volume}{88}}, \bibinfo{pages}{256601}
  (\bibinfo{year}{2002}).

\bibitem[{\citenamefont{Okamoto et~al.}(1999)\citenamefont{Okamoto, Hosoya,
  Kawaji, and Yagi}}]{Okamoto99bparal}
\bibinfo{author}{\bibfnamefont{T.}~\bibnamefont{Okamoto}},
  \bibinfo{author}{\bibfnamefont{K.}~\bibnamefont{Hosoya}},
  \bibinfo{author}{\bibfnamefont{S.}~\bibnamefont{Kawaji}}, \bibnamefont{and}
  \bibinfo{author}{\bibfnamefont{A.}~\bibnamefont{Yagi}},
  \bibinfo{journal}{Phys. Rev. Lett.} \textbf{\bibinfo{volume}{82}},
  \bibinfo{pages}{3875} (\bibinfo{year}{1999}).

\bibitem[{\citenamefont{Shashkin et~al.}(2001)\citenamefont{Shashkin,
  Kravchenko, Dolgopolov, and Klapwijk}}]{Shashkin01silicon}
\bibinfo{author}{\bibfnamefont{A.~A.} \bibnamefont{Shashkin}},
  \bibinfo{author}{\bibfnamefont{S.~V.} \bibnamefont{Kravchenko}},
  \bibinfo{author}{\bibfnamefont{V.~T.} \bibnamefont{Dolgopolov}},
  \bibnamefont{and} \bibinfo{author}{\bibfnamefont{T.~M.}
  \bibnamefont{Klapwijk}}, \bibinfo{journal}{Phys. Rev. Lett.}
  \textbf{\bibinfo{volume}{87}}, \bibinfo{pages}{086801}
  (\bibinfo{year}{2001}).

\bibitem[{\citenamefont{Pudalov et~al.}(2002)\citenamefont{Pudalov, Gershenson,
  Kojima, Butch, Dizhur, Brunthaler, Prinz, and Bauer}}]{Pudalov02silicon}
\bibinfo{author}{\bibfnamefont{V.~M.} \bibnamefont{Pudalov}},
  \bibinfo{author}{\bibfnamefont{M.~E.} \bibnamefont{Gershenson}},
  \bibinfo{author}{\bibfnamefont{H.}~\bibnamefont{Kojima}},
  \bibinfo{author}{\bibfnamefont{N.}~\bibnamefont{Butch}},
  \bibinfo{author}{\bibfnamefont{E.~M.} \bibnamefont{Dizhur}},
  \bibinfo{author}{\bibfnamefont{G.}~\bibnamefont{Brunthaler}},
  \bibinfo{author}{\bibfnamefont{A.}~\bibnamefont{Prinz}}, \bibnamefont{and}
  \bibinfo{author}{\bibfnamefont{G.}~\bibnamefont{Bauer}},
  \bibinfo{journal}{Phys. Rev. Lett.} \textbf{\bibinfo{volume}{88}},
  \bibinfo{pages}{196404} (\bibinfo{year}{2002}).

\bibitem[{\citenamefont{Tutuc et~al.}(2002)\citenamefont{Tutuc, Melinte, and
  Shayegan}}]{Tutuc02gfactor}
\bibinfo{author}{\bibfnamefont{E.}~\bibnamefont{Tutuc}},
  \bibinfo{author}{\bibfnamefont{S.}~\bibnamefont{Melinte}}, \bibnamefont{and}
  \bibinfo{author}{\bibfnamefont{M.}~\bibnamefont{Shayegan}},
  \bibinfo{journal}{Phys. Rev. Lett.} \textbf{\bibinfo{volume}{88}},
  \bibinfo{pages}{036805} (\bibinfo{year}{2002}).

\bibitem[{\citenamefont{Noh et~al.}(2002)\citenamefont{Noh, Lilly, Tsui,
  Simmons, Pfeiffer, and West}}]{Noh02gfactor}
\bibinfo{author}{\bibfnamefont{H.}~\bibnamefont{Noh}},
  \bibinfo{author}{\bibfnamefont{M.~P.} \bibnamefont{Lilly}},
  \bibinfo{author}{\bibfnamefont{D.~C.} \bibnamefont{Tsui}},
  \bibinfo{author}{\bibfnamefont{J.~A.} \bibnamefont{Simmons}},
  \bibinfo{author}{\bibfnamefont{L.~N.} \bibnamefont{Pfeiffer}},
  \bibnamefont{and} \bibinfo{author}{\bibfnamefont{K.~W.} \bibnamefont{West}}
  (\bibinfo{year}{2002}), \eprint{cond-mat/0206519}.

\bibitem[{Rev()}]{Rev01MIT}
\bibinfo{note}{For a review of the MIT, see E. Abrahams, S. V. Kravchenko and
  M. P. Sarachik, Rev. Mod. Phys., {\bf{73}}, 251 (2001)}.

\bibitem[{\citenamefont{Vitkalov et~al.}(2001)\citenamefont{Vitkalov, Zheng,
  Mertes, Sarachik, and Klapwijk}}]{Vitkalov01silicon}
\bibinfo{author}{\bibfnamefont{S.~A.} \bibnamefont{Vitkalov}},
  \bibinfo{author}{\bibfnamefont{H.}~\bibnamefont{Zheng}},
  \bibinfo{author}{\bibfnamefont{K.~M.} \bibnamefont{Mertes}},
  \bibinfo{author}{\bibfnamefont{M.~P.} \bibnamefont{Sarachik}},
  \bibnamefont{and} \bibinfo{author}{\bibfnamefont{T.~M.}
  \bibnamefont{Klapwijk}}, \bibinfo{journal}{Phys. Rev. Lett.}
  \textbf{\bibinfo{volume}{87}}, \bibinfo{pages}{086401}
  (\bibinfo{year}{2001}).

\bibitem[{\citenamefont{Shashkin et~al.}(2002)\citenamefont{Shashkin,
  Kravchenko, Dolgopolov, and Klapwijk}}]{Shashkin02mass}
\bibinfo{author}{\bibfnamefont{A.~A.} \bibnamefont{Shashkin}},
  \bibinfo{author}{\bibfnamefont{S.~V.} \bibnamefont{Kravchenko}},
  \bibinfo{author}{\bibfnamefont{V.~T.} \bibnamefont{Dolgopolov}},
  \bibnamefont{and} \bibinfo{author}{\bibfnamefont{T.~M.}
  \bibnamefont{Klapwijk}}, \bibinfo{journal}{Phys. Rev. B}
  \textbf{\bibinfo{volume}{66}}, \bibinfo{pages}{073303}
  (\bibinfo{year}{2002}).

\bibitem[{\citenamefont{Fang and Stiles}(1968)}]{Fang68tilt}
\bibinfo{author}{\bibfnamefont{F.~F.} \bibnamefont{Fang}} \bibnamefont{and}
  \bibinfo{author}{\bibfnamefont{P.~J.} \bibnamefont{Stiles}},
  \bibinfo{journal}{Phys. Rev.} \textbf{\bibinfo{volume}{174}},
  \bibinfo{pages}{823} (\bibinfo{year}{1968}).

\bibitem[{\citenamefont{Vitkalov et~al.}(2000)\citenamefont{Vitkalov, Zheng,
  Mertes, Sarachik, and Klapwijk}}]{Vitkalov00bparal}
\bibinfo{author}{\bibfnamefont{S.~A.} \bibnamefont{Vitkalov}},
  \bibinfo{author}{\bibfnamefont{H.}~\bibnamefont{Zheng}},
  \bibinfo{author}{\bibfnamefont{K.~M.} \bibnamefont{Mertes}},
  \bibinfo{author}{\bibfnamefont{M.~P.} \bibnamefont{Sarachik}},
  \bibnamefont{and} \bibinfo{author}{\bibfnamefont{T.~M.}
  \bibnamefont{Klapwijk}}, \bibinfo{journal}{Phys. Rev. Lett.}
  \textbf{\bibinfo{volume}{85}}, \bibinfo{pages}{2164} (\bibinfo{year}{2000}).

\bibitem[{\citenamefont{Dolgopolov and Gold}(2000)}]{Dolgopolov00bparal}
\bibinfo{author}{\bibfnamefont{V.~T.} \bibnamefont{Dolgopolov}}
  \bibnamefont{and} \bibinfo{author}{\bibfnamefont{A.}~\bibnamefont{Gold}},
  \bibinfo{journal}{JETP Lett.} \textbf{\bibinfo{volume}{71}},
  \bibinfo{pages}{27} (\bibinfo{year}{2000}).

\bibitem[{\citenamefont{Tutuc et~al.}(2001)\citenamefont{Tutuc, Pootere,
  Papadakis, and Shayegan}}]{Tutuc01bparal}
\bibinfo{author}{\bibfnamefont{E.}~\bibnamefont{Tutuc}},
  \bibinfo{author}{\bibfnamefont{E.~P.~D.} \bibnamefont{Pootere}},
  \bibinfo{author}{\bibfnamefont{S.~J.} \bibnamefont{Papadakis}},
  \bibnamefont{and} \bibinfo{author}{\bibfnamefont{M.}~\bibnamefont{Shayegan}},
  \bibinfo{journal}{Phys. Rev. Lett.} \textbf{\bibinfo{volume}{86}},
  \bibinfo{pages}{2858} (\bibinfo{year}{2001}).

\bibitem[{\citenamefont{Herbut}(2001)}]{Herbut01inplanefield}
\bibinfo{author}{\bibfnamefont{I.~F.} \bibnamefont{Herbut}},
  \bibinfo{journal}{Phys. Rev. B} \textbf{\bibinfo{volume}{63}},
  \bibinfo{pages}{113102} (\bibinfo{year}{2001}).

\bibitem[{not({\natexlab{a}})}]{note0.05T}
\bibinfo{note}{For i = 1, 3/2, 2, 5/2, all $B_{perp}$ is within 10\% of 0.07T.
  Data taken for i=1/2, at $B_{perp}\sim$ 0.07T and $B_{perp}$ $\sim$ 0.05T are
  identical within experimental errors.}

\bibitem[{\citenamefont{Batke and Tu}(1986)}]{Batke86massinplanefield}
\bibinfo{author}{\bibfnamefont{E.}~\bibnamefont{Batke}} \bibnamefont{and}
  \bibinfo{author}{\bibfnamefont{C.~W.} \bibnamefont{Tu}},
  \bibinfo{journal}{Phys. Rev. B} \textbf{\bibinfo{volume}{34}},
  \bibinfo{pages}{3027} (\bibinfo{year}{1986}).

\bibitem[{\citenamefont{Oelting et~al.}(1988)\citenamefont{Oelting, Wieck,
  Batke, and Merkt}}]{Batke88massinplanefield}
\bibinfo{author}{\bibfnamefont{S.}~\bibnamefont{Oelting}},
  \bibinfo{author}{\bibfnamefont{A.~D.} \bibnamefont{Wieck}},
  \bibinfo{author}{\bibfnamefont{E.}~\bibnamefont{Batke}}, \bibnamefont{and}
  \bibinfo{author}{\bibfnamefont{U.}~\bibnamefont{Merkt}},
  \bibinfo{journal}{Surf. Sci.} \textbf{\bibinfo{volume}{196}},
  \bibinfo{pages}{273} (\bibinfo{year}{1988}).

\bibitem[{not({\natexlab{b}})}]{notemg}
\bibinfo{note}{Ref.\cite{Varsano01ferro} calculates the magnetic
  susceptibility, $\sim$ m*g*$^{2}$. A direct numerical comparison is not
  available.}

\end{thebibliography}

\end{document}